\documentclass[%
 aip,
 jap,%
 amsmath,amssymb,
reprint,
nofootinbib]{revtex4-1}

\usepackage{graphicx}
\usepackage{color}
\usepackage{dcolumn}
\usepackage{bm}
\usepackage{subfigure}

\usepackage{textcomp}

\newcommand{\kacm}{\mathrm{kA}/\mathrm{cm}^2}

\usepackage{ulem}

\newcommand{\acm}{\mathrm{A}/\mathrm{cm}^2}

\begin{document}




\title{Spin-transfer torque switching below 20 kA/cm$^2$ in perpendicular magnetic tunnel junctions}

\author{Johannes Christian \surname{Leutenantsmeyer}}
\email{jleuten@gwdg.de}
\author{Marvin \surname{Walter}}
\author{Steffen \surname{Wittrock}}
\affiliation{I. Physikalisches Institut, Georg-August-Universit\"at G\"ottingen, Friedrich-Hund-Platz 1, D-37077 G\"ottingen, Germany}%
\author{Patrick~Peretzki}
\author{Henning~Schuhmann}
\author{Michael~Seibt}
\affiliation{IV. Physikalisches Institut, Georg-August-Universit\"at G\"ottingen,Friedrich-Hund-Platz 1, D-37077 G\"ottingen, Germany}%
\author{Markus \surname{M\"unzenberg}}
\affiliation{I. Physikalisches Institut, Georg-August-Universit\"at G\"ottingen, Friedrich-Hund-Platz 1, D-37077 G\"ottingen, Germany}%


\date{\today}

\begin{abstract}
We demonstrate the reduction of critical spin-transfer torque (STT) switching currents in {{Co-Fe-B}}/MgO based magnetic tunnel junctions (MTJ) with perpendicular magnetization anisotropy (PMA). 
The junctions yield tunnel magnetoresistance (TMR) ratios of up to 64\% at 4 monolayer (ML) tunnel barrier thickness. 
In this paper, the reduction of the critical switching current density is studied. By optimizing the applied bias field during DC-STT measurements, ultra low critical switching current densities of less than 20~kA/cm$^2$, even down to 9~kA/cm$^2$, are found. 
With the reduced switching currents, our samples are ideal candidates for further experimental studies such as the theoretical predicted thermally driven spin-transfer torque effect.


%
\end{abstract}


\maketitle

\section{Introduction}
Spincaloric and spintronic effects in Co-Fe-B/MgO based devices gained interest in recent research. Higher storage density, lower power consumption and faster access times are expected parameters of these devices. Current induced magnetization dynamics provide the opportunity to further enhance the storage density and performance of storage devices.\cite{Slaughter2009, Brataas2012, Walter2011}
The new field of spin-caloritronics promises to utilize excess heat for spintronic applications. One of these effects, thermal spin transfer torque (T-STT) describes magnetization reversal induced by thermally excited electron mobility. First calculations were published by Jia et al. in 2011.\cite{Jia2011} 
The effect promises the manipulation of the magnetization configuration only by applying 
a temperature gradient of the order of several Kelvin. 
We have already estimated that those temperature gradients can be achieved in our tunnel junctions by femtosecond laser excitation.
\cite{Leutenantsmeyer2013}

This paper is focussed on lowering the critical current for magnetization reversal.
A low switching current is also desired for high performance storage devices.\cite{Ikeda2010} Usual current densities of in-plane MTJs are in the range of $10^6$~$\acm$ (Ref. \onlinecite{Slaughter2009}). MTJs with a perpendicular magnetic anisotropy promise the reduction of the critical switching current while maintaining a high thermal stability $\Delta$.\cite{Ikeda2010, Wang2013}
From theory, the critical switching currents for in-plane ($I_{c}^\mathrm{I}$) and perpendicular MTJs ($I_{c}^\mathrm{P}$) are described by equations~\ref{ica} and \ref{icb}:\cite{Ikeda2010, Wang2013, Slonczewski1996, Berger1996, Sun2000}
\begin{align}
I_{c}^\mathrm{I}&= \frac{2\alpha e}{\hbar \eta} M_SV\cdot \left[ H_{k\parallel}+2\pi M_S  \right] \label{ica},
\\
I_{c}^\mathrm{P}&= \frac{2\alpha e}{\hbar \eta} M_SV \cdot H_{k_\perp}\label{icb},
\end{align}
where $\alpha$ denotes the Gilbert-damping constant, $e$ the elementary charge, $\hbar$ the reduced Planck constant, $M_S$ the saturation magnetization, $V$ the volume of the ferromagnet and $H_{k}$ the anisotropy field for in-plane and perpendicular magnetization, respectively. $\eta$ denotes the spin-torque efficiency parameter, which is depending on the spin-polarization and the relative angle beween the ferromagnets. According to Ref.~\onlinecite{Sun2000}, this parameter can be assumed to be equal to the spin-polarization $P$ for the coherent tunneling process, involving ferromagnetic electrode and barrier.
In case of in-plane MTJs, the critical current is dominated by the shape anisotropy term $2\pi M_S$. For PMA MTJs, the barrier height $E_b$, which has to be overcome to manipulate the magnetization, is $E_b=M_SH_kV/2$ and the $I_c$ is reduced.\cite{Ikeda2010, Wang2013}
Further attempts to reduce the critical switching currents exploit the application of voltage pulses to reduce the barrier height only during the magnetization reversal. 
The reported critical switching current densities by voltage induced switching are in the range of $10^4$ to $10^5$~$\acm$ (Refs. \onlinecite{Shiota2012a,Wang2013,Wang2012a}). 

The thermal stability $\Delta$ corresponds directly to the barrier height $\Delta = E_b/k_BT$. For applications, also a high thermal stability is required to retain the magnetization states for at least 10 years. This criterion corresponds to a value of $\Delta$ greater than 40.\cite{Ikeda2010} The thermal stability can be determined by equation~\ref{stabilitaet}:\cite{Ikeda2010, Sun2000, Wang2013}

\begin{align}
\Delta = \frac{E_b}{k_BT} = \frac{\eta I_c}{ \left(\frac{4e}{\hbar}\right) \alpha k_B T}=\frac{\sqrt{\frac{\mathrm{TMR}}{2+\mathrm{TMR}}} I_c}{ \left(\frac{4e}{\hbar}\right) \alpha k_B T}
\label{stabilitaet},
\end{align}
where $k_BT$ is the Boltzmann constant at room temperature and $I_c$ the averaged 
critical current for both switching directions. 

For our junctions, a Gilbert damping constant of $\alpha = 0.006(1)$ is assumed and covered by measurements on a thick Co-Fe-B film published in Ref.~\onlinecite{Ulrichs2010}. Oogane et al.\cite{Oogane2006} reported values for $\alpha$ similarly of the order of $10^{-3}$. Measurements by Iihama et al. on perpendicular \mbox{Co-Fe-B} thin films are also in good agreement. Depending on the annealing temperature values between 0.017 and 0.009 are reported.\cite{Iihama2012}
Since $\alpha$ is strongly dependent on the composition of the Co-Fe-B alloy\cite{Oogane2006}, the obtained value of the 50~nm film, sputtered from the same target, is assumed to be suitable. Though this film is significantly thicker, the obtained value is in good agreement with the 1.2~nm PMA thin film from Iihama et al.\cite{Iihama2012}


\section{Sample fabrication}
The samples are prepared on thermally oxidized silicon substrates.
The grown MTJ stack consists of Ta (15~nm) / Co-Fe-B (1.0~nm) / MgO (0.84~nm) / Co-Fe-B 
(1.2~nm) / Ta (5.0~nm) / Ru (3.0~nm). 
Tantalum and Co-Fe-B are magnetron sputtered in a chamber at base pressures in the range 
of approx. $5\cdot 10^{-10}$ mbar. The MgO barrier and ruthenium capping layer are 
e-beam evaporated in an interconnected chamber at pressures below $5\cdot 10^{-10}$ 
mbar. To crystallize the Co-Fe-B electrodes, the prepared samples are annealed for 60 
minutes at 300$^\circ$C.

After annealing, the samples are patterned using standard UV- and electron-beam lithography as well as argon ion milling techniques. The junctions are of circular shape with diameters between 100~nm and 250~nm.
TMR and current-voltage (IV) characteristics are measured in two-terminal geometry at room temperature. The STT data (shown in Fig.~\ref{005Macm}) is obtained from the IV curves.

\section{Results and Discussion}
\begin{figure}[tb]
\centering
\includegraphics[width=1\linewidth]{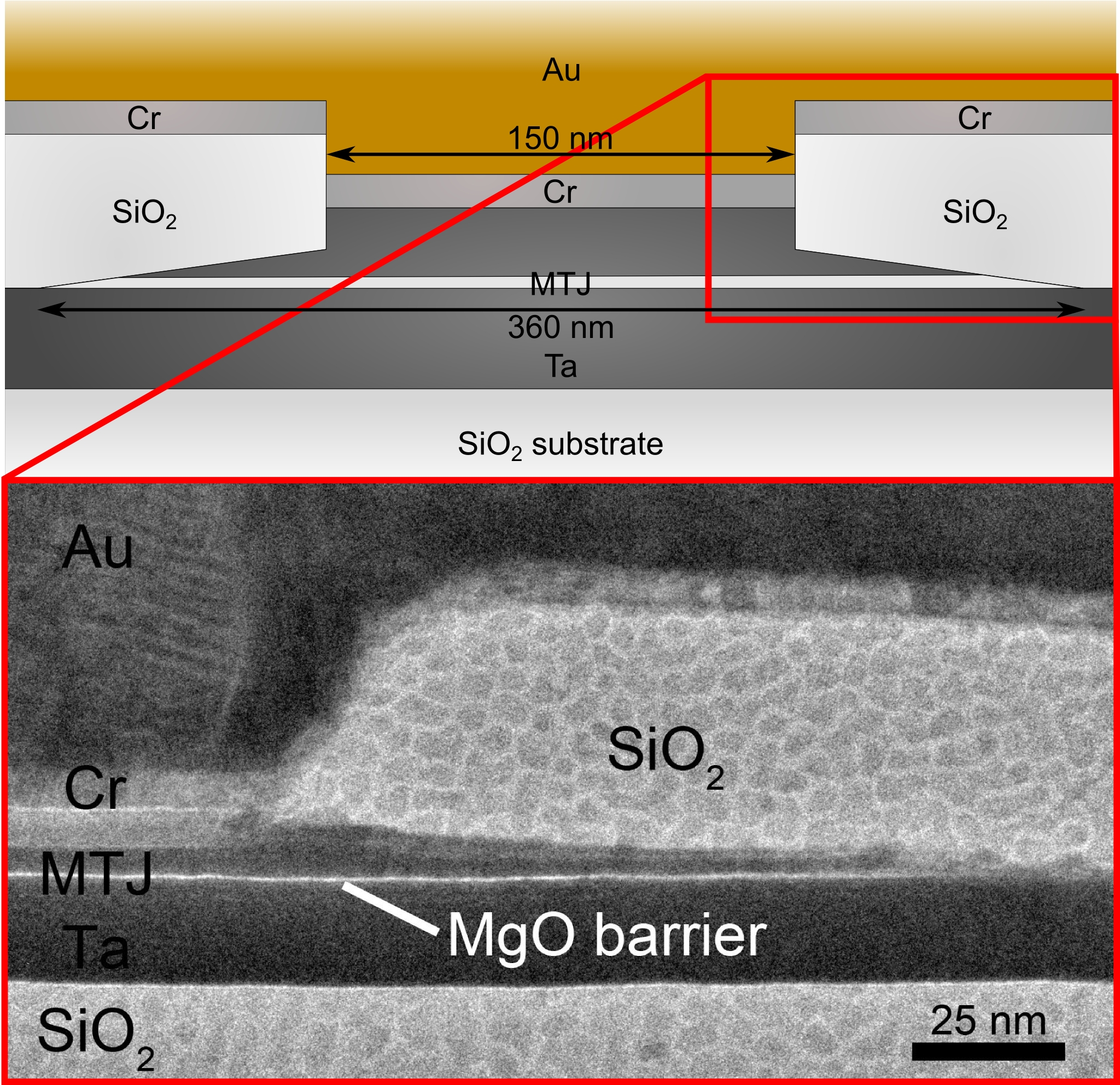}
\caption{TEM image of the sidewall of a patterned junction. The diameter of 150~nm at the top of the junction is increased to 360~nm at the position of the MgO barrier. The current is injected predominately at the center of the junction.}
\label{TEMFlanke}
\end{figure}

\begin{figure*}[tb]
\centering
\subfigure[]{\includegraphics[width=0.49\textwidth]{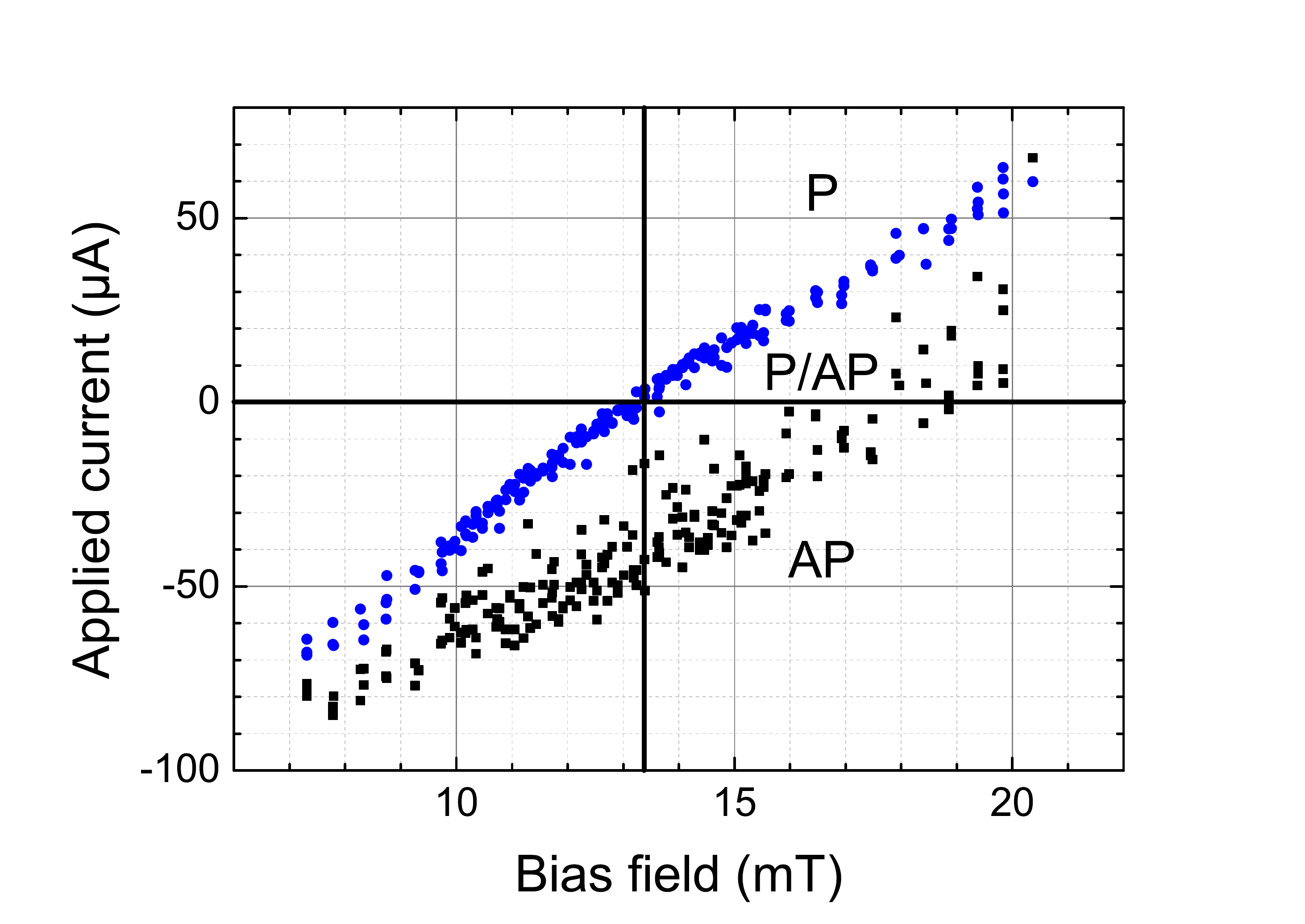}}
\subfigure[]{\includegraphics[width=0.49\textwidth]{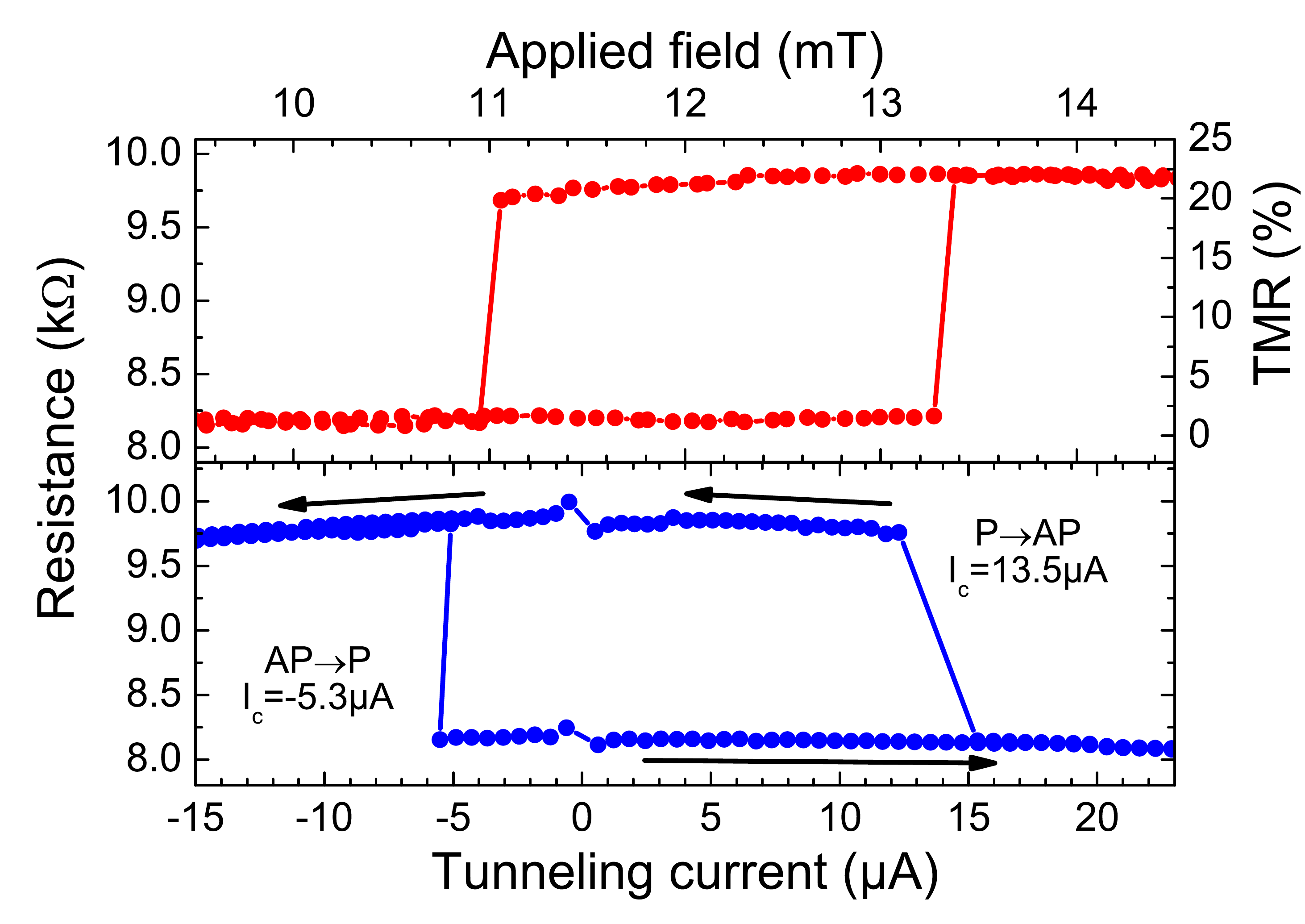}}
\caption{(a) The switching phase diagram of a 360~nm PMA MTJ. The position of the minor loop is marked by the horizontal, the STT data is marked by the vertical black line. (b) Electrical characterization of the junction. The minor loop (top) yields a TMR ratio of 22\%. At a bias field of 13.4~mT the average of both critical switching currents (9.4~\textmu A) equals a critical current density of only $9 \pm 2$~$\kacm$ (bottom).}
\label{005Macm}
\end{figure*}

First, the magnetoresistive behavior of the junctions is characterized. 
We find TMR ratios for PMA MTJs with barrier thickness from 3--4 monolayers of up to 55--64\%.\cite{Leutenantsmeyer2013} Here we discuss a junction, where the smallest $J_c$ was realized.
The magnetic minor loop (Fig.~\ref{005Macm}b) exhibits clear PMA. The obtained TMR ratio is 22\%.
To calculate the resistance area product, the junction size is determined with TEM 
imaging. Due to the slope of the sidewall (see Fig.~\ref{TEMFlanke}), resulting from the 
argon ion milling process, the radius of the MgO barrier is increased from nominal 75~nm 
at the top of the junction to 180~nm. Using the first, upper diameter, the 
resistance-area product is 143~$\Omega$\textmu m$^2$. Using the junction diameter, 
including the sidewalls extracted from the bottom of the cross section, 
823~$\Omega$\textmu m$^2$ are found for the 4 monolayer thick MgO barrier. 
To detect the critical currents for spin-transfer torque switching, the IV characteristics are recorded while the applied bias field is varied. From the data, the critical currents are extracted and both values averaged. The critical current density ($J_c$) is calculated using 360~nm as junction diameter.


Fig.~\ref{005Macm}a contains the switching phase diagram. 
For bias fields between 7.3~mT and 20.4~mT, STT switching is observed. Minimal switching currents are found between 13~mT and 18~mT, in average $19 \pm 5$~$\kacm$. As of today, values in this range were only reported by voltage induced switching.


Fig.~\ref{005Macm}b depicts the electrical characterization of the MTJ. The switching between parallel and antiparallel alignment occurs at critical currents of $-5.3 \pm 0.3$~\textmu A and $13.5 \pm 0.3$~\textmu A at a bias field of 13.4~mT. The minimal averaged critical current density corresponds to a value as low as $9 \pm 2$~$\kacm$ (360~nm in diameter, 0.1~\textmu m$^2$ at $9.4 \pm 1.5$~\textmu A). This value is significantly lower than any published $J_c$ for DC-STT measurement in PMA MTJs so far.\cite{Ikeda2010, Sato2013, Meng2011b} Assuming the smaller diameter of 150~nm, in case of an inhomogeneous current distribution, $J_c$ would still be on the order of $53 \pm 2$~$\kacm$.

The critical torque $\tau_c$ can be estimated from the measured $J_c$, as given in 
Ref.~\onlinecite{Slonczewski2005}:
\begin{align}
\tau_c = \frac{\hbar \eta}{2e}J_c = \frac{\hbar 
\sqrt{\frac{\mathrm{TMR}}{2+\mathrm{TMR}}}}{2e}J_c \label{casimirstolleformel}.
\end{align}
Using the experimentally observed averaged $J_c$ of 19~$\kacm$ and a TMR ratio of 22\%, 
a threshold torque of 19 nJ/cm$^2$ is obtained. The thermal torques per Kelvin 
temperature difference calculated by Jia et al. ranged from 3.3~nJ/m$^2$ for a 5 
monolayer to 195~nJ/m$^2$ for a 3 monolayer MgO barrier.
Thus, our experimentally obtained critical torque is small enough for T-STT at temperature gradients of a few Kelvin.

The thermal stability criterion for storage applications is, however, not yet matched ($\Delta = 19.6 \pm 0.7$ for $\alpha  = 0.006$), mainly due to the relatively low TMR ratio. Isogami et al. reported increased TMR ratios after in-situ heat-treatment of a 4 monolayer MgO MTJ to over 200\%. The application of heat is reported to reduce the number of grain boundaries in the solid state epitaxy process.\cite{Isogami2008} In our junctions, the texture of the MgO barrier was found to be enhanced by in-situ heating.\cite{Leutenantsmeyer2013} According to Ref.~\onlinecite{Isogami2008}, in a columnar growth model less boundaries improve the coherent tunneling process, increase the TMR ratio and lower the resistance-area product. 
It is clear that for a T-STT-MRAM device the TMR ratio of our MTJs should be raised to 200\% similarly. Then the actual critical currents would lead to a $\Delta$ of $44.0 \pm 1.4$, which already satisfies the thermal stability requirement.

\section{Conclusion}
We have demonstrated PMA MTJs with ultra thin MgO barriers, a TMR ratio of 22\% and spin-transfer torque switching. 
Bias field dependent DC-STT measurements revealed threshold currents for magnetization reversal of less than 20~$\kacm$. A critical current density of $J_c = 9 \pm 2$~$\kacm$ is the lowest reported value so far. The derived torques are in the expected range to observe T-STT. 
Further optimization is needed to fulfill the thermal stability requirements.

\begin{acknowledgments}
The authors acknowledge funding from the German Research Foundation (DFG) through 
SPP~1538 \textit{``Spin Caloric Transport''}. 
\end{acknowledgments}


\bibliography{bib}

\end{document}